\documentstyle[eqsecnum,twocolumn,aps]{revtex}

\newcommand{\ch}{\v{c}}

\newcommand{\beq}{\begin{equation}}
\newcommand{\eeq}{\end{equation}}
\newcommand{\bdm}{\begin{displaymath}}
\newcommand{\edm}{\end{displaymath}}
\newcommand{\beqa}{\begin{eqnarray}}
\newcommand{\eeqa}{\end{eqnarray}}
\newcommand{\beqab}{\begin{eqnarray*}}
\newcommand{\eeqab}{\end{eqnarray*}}
\def\nn{\nonumber}

\begin{document}
\draft
\preprint{}
\title{XXZ spin chain in transverse field as a regularization of the
             sine-Gordon model}  
\author{Silvio Pallua and Predrag Prester}
\address{Department of Theoretical Physics, University of Zagreb\\
Bijeni\ch ka c.32, POB 162, 10001 Zagreb, Croatia}
\date{\today}
\maketitle
\begin{abstract}
We consider here XXZ spin chain perturbed by the operator $\sigma^{x}$
(``in transverse field'') which is a lattice regularization of the
sine-Gordon model. This can be shown using conformal perturbation theory.
We calculated mass ratios of particles which lie in a discrete part of
the spectrum and obtained results in accord with the DHN formula and in 
disagreement with recent calculations in literature based on numerical Bethe
Ansatz and infinite momentum frame methods. We also analysed a short
distance behavior of this states (UV or conformal limit). Our result for
conformal dimension of the second breather state is different from the
conjecture in [Klassen and Melzer, Int. J. Mod. Phys. A {\bf 8}, 4131 (1993)]
and is consistent whith this paper for other states.
\end{abstract}

\narrowtext

\section{Introduction}
\label{sec:intro}

The sine-Gordon (SGM) and massive Thirring (MTM) models in two dimensions
belong to a group of the most studied QFT's and are certainly the best
understood non-trivial massive field theories. A large number of different
techniques have been successfully tested on these models and they led us
to a number of interesting results, including the famous duality relation
between them \cite{Col,Man,DeGW}.

Regarding a mass spectrum, we can classify all methods in basicaly three
groups: (a) the semi-classical DHN method \cite{DHN}, (b) factorized scattering
theory \cite{Zam1} and (c) methods based on Bethe Ansatz, which can be further
subdivided in continuum ones \cite{BeTh,Kor} and in discrete ones
\cite{Lut,Lusch} (some lattice regularizations were used). Results of all
methods were the same; beside soliton and antisoliton (fermion and antifermion
in MTM language) there are bound states (breathers) and their masses are given
by:
\beq \label{dhn}
m_{n}=2m\sin\frac{n\pi\beta^{2}}{2(8\pi-\beta^{2})},\quad n=1,2,\ldots <
\frac{8\pi}{\beta^{2}}-1
\eeq
where $m$ is the soliton mass and $\beta$ is the coupling constant in SGM (see
(\ref{lsg})). Because of the Coleman's theorem of equivalence between the 
SGM and the MTM in soliton number (charge) zero sector (proved using
perturbative expansion in mass), the same spectrum should be valid for the
MTM. Using standard conventions (as in \cite{Col}), a connection (``duality
relation'') between $\beta$ and the MTM coupling constant $g_{0}$ (in the
Schwinger normalization) is given by:
\bdm 
1+\frac{g_{0}}{\pi}=\frac{4\pi}{\beta^{2}}
\edm

However, recently appeared claims \cite{FuOg,FuSeYa,FuHi} that mass spectrum of
the MTM is different than (\ref{dhn}) and that there is only one breather in
the whole interval $g_{0}>0$ (for negative values of $g_{0}$ fermion and
antifermion repel each other and there are no bound states, like in
(\ref{dhn})). In \cite{FuOg}, using the infinite momentum frame technique
and working only in $q\bar{q}$ sector of the Fock space (neglecting
$qq\bar{q}\bar{q}$ and higher fermion components), authors obtained the mass
of the (only) breather:
\beq \label{fum}
M=2m\cos\alpha
\eeq
where the parameter $0\le\alpha<\pi/2$ is obtained by solving equation:
\bdm 
\frac{\tan\alpha}{\frac{\pi}{2}-\alpha}=\frac{g}{\pi}\left[1+\frac{1}{\cos^{2}
\alpha}\left(1-\frac{g}{4\pi}\right)\right]
\edm
and $g$ is the MTM coupling constant in Johnson's normalization which is
connected to that in Schwinger's normalization by:
\bdm 
g_{0}=\frac{2g}{2-\frac{g}{\pi}}
\edm

Afterwards, in \cite{FuSeYa} authors reexamined an analysis of \cite{BeTh},
but contrary to \cite{BeTh} they numericaly solved Bethe Ansatz equations
for a finite space extension and a finite number of quasi-particles, and
after that made an extrapolation to infinity. Their analysis confirmed
results of \cite{FuOg}; they found only one breather, with the mass in 
good agreement with (\ref{fum}).

In this paper we propose ourselves to calculate certain properties of the
SGM like mass ratios and scaling dimensions of operators creating particle
states.
Using the conformal perturbation theory \cite{Zam2,KlMelz} it can be
shown that XXZ spin chain with an even number of sites and periodic boundary
conditions in a transverse magnetic field ($\sigma^{x}$ perturbation) is spin
chain regularization of the SGM (see Appendix B in \cite{KlMelz}). We numericaly
diagonalize the spin chain Hamiltonian up to 16 sites and extrapolate results
to infinite lenght continuum limit using the BST extrapolation algorithm
\cite{BulSt,HenSchutz}. The same method was previously applied to conformal
unitary models perturbed by some relevant (usually termal) operator
\cite{HenSal,Hen,Hen2}. In this way we can obtain estimates of mass ratios
without further assumptions, particularly those criticized in
\cite{FuOg,FuSeYa,FuHi}.

Results of our analysis are as follows. For a whole range of the coupling
constants we can cover ($0<\beta\le\sqrt{2\pi}$) our results agree with the
DHN formula (\ref{dhn}) and disagree with (\ref{fum}), i.e. results of
\cite{FuOg,FuSeYa}. Of course, we couldn't say anything about breathers
higher then third because they lie in a continuum part of the spectrum
($m_{n}>2m_{1}$ for $n\ge 3$). We should
also say that precision in this method is far from that achieved by, e.g.
Bethe Ansatz methods, so we can't claim that DHN formula is exact.

Finaly, as a by product, we studied UV limit of particle states. It agrees with
that conjectured in \cite{KlMelz} for (anti)soliton and first breather. However,
for the second breather we obtain the same scaling dimension as for the first,
contrary to \cite{KlMelz}.

\section{The SGM as a massive perturbation of the Gaussian model}
\label{sec:gauss}

The SGM is a $1+1$ dimensional field theory of a pseudoscalar field $\varphi$,
defined classically by the Lagrangian:
\beq \label{lsg}
{\cal L}_{SG}=\frac{1}{2}\partial_{\mu}\varphi\partial^{\mu}\varphi+\lambda
\cos(\beta\varphi)
\eeq
Here $\lambda$ is a mass scale (with mass dimension depending on a
regularization scheme), $\beta$ is a dimensionless coupling (which does not
renormalize) and one identifies field configurations that differ by a period
$2\pi/\beta$ of the potential (because we want to have ``ordinary'' QFT with
a unique vacuum).

In \cite{KlMelz} it was shown that SGM can be viewed as a perturbed CFT when
second term in (\ref{lsg}) is treated as a (massive) perturbation. We'll now
repeat here relevant results of their analyses.

An unperturbed theory $\lambda=0$ (approached in UV limit) is the free massless
compactified pseudoskalar CFT (known as Gaussian model). It is conventional
to use $\Phi\equiv\sqrt{\pi}\varphi$, so that the radius of compactification
$r$, defined by equivalence $\Phi\sim\Phi+2\pi r$ is connected to $\beta$ with
\beq \label{rbe}
r=\frac{\sqrt{\pi}}{\beta}
\eeq
Solution of the equation of motion in Euclidean space, $\partial\bar{\partial}
\Phi(z,\bar{z})=0$, is
\bdm 
\Phi(z,\bar{z})=\frac{1}{2}(\phi+\bar{\phi})
\edm
The Gaussian model is a CFT with central charge $c=1$ and an operator algebra
generated by the primary fields $V_{m,n}$
\beq \label{vmn}
V_{m,n}=\: :e^{i\frac{m}{r}\Phi(z,\bar{z})+i2nr\tilde{\Phi}(z,\bar{z})}:
\eeq
where $\tilde{\Phi}\equiv(\phi-\bar{\phi})/2$. Conformal dimensions of
$V_{m,n}$ are
\beq \label{cdv}
(\Delta_{m,n},\bar{\Delta}_{m,n})=\left(\frac{1}{2}\left(\frac{m}{2r}+nr\right)
^{2},\frac{1}{2}\left(\frac{m}{2r}-nr\right)^{2}\right)
\eeq
so that it's scaling dimension and (Lorentz) spin are:
\beqab 
d_{m,n}&=&\Delta_{m,n}+\bar{\Delta}_{m,n}=\left(\frac{m}{2r}\right)^{2}+
(nr)^{2}=\frac{m^{2}\beta^{2}}{4\pi}+\frac{n^{2}\pi}{\beta^{2}} \\
s_{m,n}&=&\Delta_{m,n}-\bar{\Delta}_{m,n}=mn 
\eeqab
It is understood that $V_{m,n}$ are normalized so that
\bdm 
\langle V_{m,n}(z,\bar{z})V_{m,n}(0,0)\rangle = \delta_{m,-m'}\delta_{m,-m'}
z^{-2\Delta_{m,n}}\bar{z}^{-2\bar{\Delta}_{m,n}}
\edm
Because of $V^{\dagger}_{m,n}=V_{-m,-n}$, we can define hermitian combinations
\beqab
V^{(+)}_{m,n}\equiv\frac{1}{2}\left( V_{m,n}+V_{-m,-n}\right) \\
V^{(-)}_{m,n}\equiv\frac{i}{2}\left( V_{-m,-n}-V_{m,n}\right)
\eeqab
which will be usefull later.

In \cite{KlMelz} it is argued that an UV limit of the SGM is generated by
\beq \label{lbg}
L_{b}=\left\{ V_{m,n}|m,n\in\mathbf{Z}\right\}
\eeq
We suppose that Hilbert space of the full (perturbed) theory is isomorphic to
that of the unperturbed theory. From (\ref{rbe}) and (\ref{vmn}) follows that
a (properly normalized) perturbing operator in the SGM (\ref{lsg}) is
\beq \label{per}
\cos(\beta\varphi)=V^{(+)}_{1,0}
\eeq
which means that $\lambda$ has mass dimension $y=2-d_{1,0}=2-\frac{\beta^2}{4
\pi}$. From the condition of relevancy of the perturbation, i.e.~$y>0$, we
obtain Coleman's bound $\beta^2<8\pi$. Also, from (\ref{lbg}) and (\ref{per})
we can see that SGM has $\tilde{U}(1)\times Z_2\times\tilde{Z}_2$ internal
symmetry group. The $\tilde{U}(1)$ acts as shift on $\tilde{\Phi}$, i.e.~$
V_{m,n}\rightarrow e^{i\alpha n}V_{m,n}$, while $Z_2$ and $\tilde{Z}_2$ are
generated by $R:(\Phi,\tilde{\Phi})\rightarrow(-\Phi,\tilde{\Phi})$ (i.e.~$
V_{m,n}\rightarrow V_{-m,n}$) and $\tilde{R}:(\Phi,\tilde{\Phi})\rightarrow
(\Phi,-\tilde{\Phi})$ (i.e.~$V_{m,n}\rightarrow V_{m,-n}$), respectively.

To conclude this section, consider the SGM defined on a cilinder with
infinite time dimension and space extension equal to $L$. There are three
independent constants with which we can express all quantities in the theory,
$\beta$, $\lambda$ and $L$ with mass dimensions $d_{\beta}=0$, $d_{\lambda}=
2-d_{1,0}=2-\beta^{2}/(4\pi)$ and $d_{L}=-1$. It is usefull to define
dimensionless scaling parameter $\mu$
\beq \label{msp}
\mu\equiv\lambda L^{d_{\lambda}}=\lambda L^{2-\frac{\beta^{2}}{4\pi}}
\eeq
and use $\beta$, $\mu$ and $\lambda$ as a set of independent parameters. Now,
from ordinary dimensional analysis follows that any quantity $X$ in the
theory, with mass dimension $d_{X}$, can be written as
\beq \label{xlg}
X=\lambda^{\frac{d_{X}}{d_{\lambda}}}g_{X}(\beta,\mu)=
\lambda^{\frac{d_{X}}{2-\frac{\beta^{2}}{4\pi}}}g_{X}(\beta,\mu)
\eeq
where $g_{X}$ is the scaling function connected to $X$. We see that all
dimensionless quantities depend only on $\beta$ and $\mu$. Especially, we have
for masses of particles:
\beq \label{mlg}
m_{i}(\beta,\mu,\lambda)=\lambda^{(2-\frac{\beta^{2}}{4\pi})^{-1}}G_{i}(\beta,
\mu)
\eeq
Now, there are two interesting limits. The first one is the infinite lenght
limit, $L\to\infty$, which is equal to $\mu\to\infty$ (see (\ref{msp})).
We are interested here in mass ratios:
\bdm 
r_{i}(\beta)=\lim_{\mu\to\infty}\frac{m_{i+1}(\beta,\mu,\lambda)}{m_{1}(\beta,
\mu,\lambda)}=\lim_{\mu\to\infty}\frac{G_{i+1}(\beta,\mu)}{G_{1}(\beta,\mu)}
\edm

The second interesting limit is the UV limit given by $L\to 0$ ($\mu\to 0$).
Basic assumption of conformal perturbation theory is that the perturbed
QFT should aproach smoothly to CFT in UV limit. It means that if we write
(\ref{xlg}) in the form:
\beq \label{xcp}
X=X_{CFT}(\beta,L)+\lambda^{\frac{d_{X}}{d_{\lambda}}}h_{X}(\beta,\mu)
\eeq
where $X_{CFT}$ is the value for $X$ in conformal point ($\lambda=0$), than a
Taylor expansion for $\mu^{d_{X}/d_{\lambda}}h_{X}(\beta,\mu)$ around $\mu=0$
will have finite radius of convergence and $h_{X}(\beta,0)=0$. Specially, for
the mass gaps we have well-known formula:
\bdm 
(m_{i})_{CFT}=\frac{2\pi}{L}d_{i}
\edm
where $d_{i}$ is the scaling dimension of the operator which creates that state
from the vacuum. Now from (\ref{mlg}), (\ref{xcp}) and (\ref{msp}) follows:
\beqa
m_{i}(\beta,\mu,\lambda)&=&\frac{2\pi}{L}d_{i}+\lambda^{\frac{1}{d_{\lambda}}}
H_{i}(\beta,\mu) \nn \\ &=& \lambda^{\frac{1}{d_{\lambda}}}\left[ 2\pi d_{i}
\mu^{-\frac{1}{d_{\lambda}}}+H_{i}(\beta,\mu)\right] \label{mch} \\ 
&=& \lambda^{(2-\frac{\beta^{2}}{4\pi})^{-1}}\left[ 2\pi d_{i}\mu^{-(2-
\frac{\beta^{2}}{4\pi})^{-1}}+H_{i}(\beta,\mu)\right] \nn
\eeqa
Now, what are scaling dimensions of zero-momentum one-particle states in SGM,
i.e.~of soliton, antisoliton and breathers? In Table \ref{klme} we show values
conjectured in \cite{KlMelz}. In Sec.~\ref{sec:conf} we'll show that we
obtain a different result for the second breather.

\section{Spin chain regularization of the SGM}
\label{sec:chain}

It was proposed (Appendix B in \cite{KlMelz} ) that XXZ spin chain with
periodic boundary conditions in a transverse magnetic field defined by the
Hamiltonian:
\beqa
H=-\sum_{n=1}^{N}\left( \sigma_{n}^{x}\sigma_{n+1}^{x}+\sigma_{n}^{y}
\sigma_{n+1}^{y}+\Delta\sigma_{n}^{z}\sigma_{n+1}^{z}+h\sigma_{n}^{x}\right)
\nn \\ \vec{\sigma}_{N+1}\equiv\vec{\sigma}_{1} \hspace{2cm} \label{htf}
\eeqa
where $\sigma_a$ are Pauli matrices, $N$ is an even integer, $-1\le\Delta <1$
(we use the usual parametrization $\Delta=-\cos\gamma$, $0\le\gamma <\pi$), is
a spin chain regularization of the SGM. The argument has two steps; first one
must show that unperturbed theories are equivalent, i.e.~that (\ref{htf}) with
$h=0$ is a spin chain regularization of $L_{b}$ CFT (\ref{lbg}), and second
that in the unperturbed theory ($h=0$) perturbation operator $\sigma_{n}^{x}$
is a lattice regularization of $V_{1,0}^{(+)}(x)$.

For a first step one must take $h=0$ in (\ref{htf}), i.e.~to
consider periodic XXZ spin chain
\beqa
H_{XXZ}=-\sum_{n=1}^{N}\left(\sigma_{n}^{x}\sigma_{n+1}^{x}+\sigma_{n}^{y}
\sigma_{n+1}^{y}+\Delta\sigma_{n}^{z}\sigma_{n+1}^{z}\right)
\nn \\ \vec{\sigma}_{N+1}\equiv\vec{\sigma}_{1} \hspace{2cm} \label{xxz}
\eeqa
$H_{XXZ}$ commutes with $S^{z}=1/2\sum_{n=1}^{N}
\sigma_{n}^{z}$. We denote eigenvalues of $S^{z}$ by $Q$. $Q$ is integer
(half-odd integer) when $N$ is even (odd) and $-N/2\le Q\le N/2$. $H_{XXZ}$ is
also translation-invariant where translations by one site are generated by:
\beq \label{txz}
T=\prod_{n=1}^{\stackrel{\scriptstyle \frown}{N-1}}
\frac{1}{2}\left(\vec{\sigma}_{n}\cdot\vec{\sigma}_{n+1}+1\right)
\eeq
and we define (lattice) momentum operator by $T=\exp(-iP)$. From (\ref{txz})
follows that $T^{N}=1$, so eigenvalues $P_{k}$ of the
lattice momentum $P$ are given by:
\beq \label{pxz}
P_{k}=\frac{2\pi}{N}k,\quad k=0,1,\ldots,N-1
\eeq
Obviously, $P_{k}$ are defined $\bmod \,2\pi$.

Now, in \cite{Alc1} it has been shown that energy-momentum spectrum of the
periodic XXZ chain in charge sector $Q$ has following
asimptotic form for large $N$:
\begin{mathletters} \label{eptl}
\beqa \label{etl}
E_{Q,\nu}^{n,\bar{n}}&=&Ne_{\infty}+\frac{2\pi\zeta}{N}\left(\Delta_{Q,\nu}^{n}
+\bar{\Delta}_{Q,\nu}^{\bar{n}}-\frac{c}{12}\right)
\\ P_{Q,\nu}^{l}&=&\frac{2\pi}{N}\left(\Delta_{Q,\nu}^{n}-
\bar{\Delta}_{Q,\nu}^{\bar{n}}\right)+\pi\kappa_{Q,\nu}
\label{ptl} \eeqa \end{mathletters}
where $\nu\in\mathbf{Z}$, $n,\bar{n}\in\mathbf{N}^0$, central charge $c=1$,
$\kappa_{Q,\nu}\in\{0,1\}$, and conformal dimensions $\Delta_{Q,\nu}^{n}$
and $\bar{\Delta}_{Q,\nu}^{\bar{n}}$ are given by
\beq \label{cdu}
\left(\Delta_{Q,\nu}^{n},\bar{\Delta}_{Q,\nu}^{\bar{n}}\right)=\left(
\frac{1}{2}\left[\frac{Q}{2r}+r\nu\right]^{2}+n,\frac{1}{2}\left[
\frac{Q}{2r}-r\nu\right]^{2}+\bar{n}\right)
\eeq
where compactification radius is $r=[2(1-\gamma/\pi)]^{-1/2}$. From
(\ref{eptl}) and (\ref{cdu})
we can infer that continuum limit of $H_{XXZ}$ defined by:
\begin{mathletters} 
\beqa \label{chu}
H_{XXZ}^{cont}&\equiv&\frac{1}{\zeta}
\lim_{\stackrel{N\to\infty}{\scriptscriptstyle a\to 0}}
\frac{1}{a}(H_{XXZ}-Ne_{\infty}) \\
P^{cont}&\equiv&\lim_{\stackrel{N\to\infty}{\scriptscriptstyle a\to 0}}
\frac{1}{a}(P-\pi\kappa)
\label{cpu} \eeqa \end{mathletters}
($a$ is lattice constant and $L=Na$ is kept fixed) defines $c=1$ CFT, and in
fact contains $L_{b}$ of the Gaussian model as we shall see. In (\ref{cpu})
$\kappa$ is an operator which project states having ``nonuniversal
macroscopic momentum'' equal to $\pi$ (see \cite{GrimSch}). We shall comment
more on this at the end of this section. $\zeta$ is a normalization factor and
$e_{\infty}$ is (c-number) nonuniversal bulk energy density. Nonuniversal
quantities are subtracted in QFT limit.

Let us see how one can obtain $L_{b}$ and $L_{f}$ from $H_{XXZ}^{cont}$. First,
from (\ref{cdu}) and it is obvious that
\bdm 
\left(\Delta_{Q,\nu}^{0},\bar{\Delta}_{Q,\nu}^{0}\right)=\left(
\Delta_{Q,\nu},\bar{\Delta}_{Q,\nu}\right)
\edm
where $\Delta_{m,n}$ and $\bar{\Delta}_{m,n}$ are conformal dimensions
(\ref{cdv}) of the
vertex operator $V_{m,n}$ in Gaussian model. Comparing (\ref{cdu}) with
(\ref{lbg}), it is obvious that $Q$ must be integer, so $N$ must be even, and
\beq \label{leh}
L_{b}\left( r=\left[ 2\left( 1-\frac{\gamma}{\pi}\right)\right]^{-\frac{1}{2}}
\right)\iff H_{XXZ}^{cont}(\gamma)
\eeq

So, in (\ref{leh}) is given the  first half of equivalence between (\ref{htf})
and the SGM, that unperturbed CFT's are equivalent. Now one must show the
second part, that operator $\sigma_{n}^{x}$ is lattice counterpart of
$V_{1,0}^{(+)}(x)$ ($x=na$) in the Gaussian model. In \cite{LutPesch} it was
shown (in the leading order in the lattice constant $a$) that:
\beq \label{sev}
\sigma_{n}^{\pm}\propto a^{d_{1,0}}V_{\pm 1,0}(x)=a^{\frac{\beta^{2}}{4\pi}}
V_{\pm 1,0}(x)
\eeq
where $x=na$. The constant of proportionality in (\ref{sev}) is in fact known
\cite{LukZam,Luk} but we'll not need it here. So, from (\ref{sev}) we see that:
\beq \label{sxev}
\sigma_{n}^{x}\propto V^{(+)}_{1,0}(x)\qquad x=na
\eeq
in the leading order. That finally completes the argument \cite{KlMelz} that
Hamiltonian (\ref{htf}) is a spin chain regularization of the SGM where
connection between coupling constants is
\beq \label{brg}
\beta =\frac{\sqrt{\pi}}{r}=\sqrt{2(\pi-\gamma)}
\eeq
Let us made a comment on internal symmetries of continuum and lattice models.
As we emphasized in the last section SGM posesses $Z_{2}\times\tilde{Z}_{2}
\times\tilde{U}(1)$ symmetry and is integrable. But spin chain (\ref{htf}) is
only symmetric on $Z_{2}$ generated by ``charge conjugation operator'' $C$:
\bdm 
C=\prod_{n=1}^{N}\sigma_{n}^{x}
\edm
and in fact is believed to be non-integrable. That spin chain representation
of a QFT has less symmetries is not something new \cite{HenSal}.

Now, what are the relations between dimensionfull parameters ($L$, $\lambda$,
$\mu$) in the (continuum) SGM and parameters ($N$, $h$) in (lattice)
(\ref{htf}). From (\ref{chu}) and (\ref{leh}) follows
\bdm 
H_{SGM}(L)=\frac{1}{\zeta}\lim_{\stackrel{N\to\infty}{\scriptscriptstyle
a=L/N}}\frac{H}{a}
\edm
So, if we denote by $\tilde{m}_{i}$ mass gaps in the spin chain, we have:
\beq \label{mem}
m_{i}(L)=\frac{1}{\zeta}\lim_{\stackrel{N\to\infty}{\scriptscriptstyle a=L/N}}
\frac{\tilde{m}_{i}}{a}
\eeq
Also, from (\ref{sxev}) we have:
\beq \label{hla}
h\propto\lim_{a\to 0}\lambda a^{d_{\lambda}}=\lim_{a\to 0}\lambda
a^{2-\frac{\beta^{2}}{4\pi}}
\eeq
where factor of proportionality is finite. Of course, we have $L=Na$ and
$\lambda$ fixed. We can see from (\ref{hla}) that $h\to 0$ because
$d_{\lambda}>0$.
We can now express scaling parameter $\mu$ using lattice constants:
\beq \label{mhn}
\mu=\lambda L^{d_{\lambda}}\propto
\!\lim_{\stackrel{N\to\infty}{\scriptscriptstyle L,\lambda\;\mathrm{finite}}}
\!\! hN^{d_{\lambda}}
\eeq
Constant of proportionality is not important for us because we are interested
here only in $L\to\infty$ ($\mu\to\infty$) and $L\to 0$ ($\mu\to 0$) limits.
If we define now
\beq \label{tmh}
\tilde{\mu}\equiv hN^{d_{\lambda}}=hN^{2-\frac{\beta^{2}}{4\pi}}=
hN^{\frac{3}{2}+\frac{\gamma}{2\pi}}
\eeq
from (\ref{mlg}), (\ref{mem}), (\ref{mhn}) and (\ref{tmh}) we can see that:
\beq \label{tmhg}
\tilde{m}_{i}=h^{(2-\frac{\beta^{2}}{4\pi})^{-1}}\tilde{G}_{i}(\gamma,
\tilde{\mu})
\eeq
where $\gamma$ is connected to $\beta$ by (\ref{brg}). Strictly speaking,
scaling law (\ref{tmh}) should be exactly valid only in the continuum limit
$N\to\infty$, $a\to 0$ and $h\to 0$ where $L=Na$ and $\lambda\propto
ha^{\frac{\beta^{2}}{4\pi}-2}$ are kept fixed. For finite $N$ (\ref{tmh}) is
only aproximate and we expect that scaling is worser for $N$ smaller.

\begin{center} *\ \ *\ \ * \end{center}

To keep our promise, we shall now comment subtraction of ``nonuniversal
momentum'' $\pi$ mentioned in the part of the text following eq.~(\ref{cpu}),
which doesn't sound very natural (maybe ''too statistical''). A more natural
explanation is based on the fact that SGM is equivalent to (\ref{htf}) when
number of lattice sites $N$ is {\em even}. Let's suppose that lattice is
staggered, i.e.~that (in continuum limit terms) real space translations are
given by translations by {\em even} number of sites, and translation by one
site is some internal state transformation \cite{foot1}. A consequence
is that $T^{2}$ is the ``real'' lattice translation operator, so $2P$ is the
``real'' momentum which is also defined $\bmod 2\pi$. But, now we must
multiply (\ref{ptl}) by $2$, so how can we obtain the same conformal dimensions
$\Delta$ and $\bar{\Delta}$. An explanation is that the continuum spatial
extension of the system is now $L=aN/2$, so we must put $N/2$ in place of $N$
in (\ref{etl}). In (\ref{ptl}) it just compensates
factor $2$, and in (\ref{etl}) we already needed scaling factor $\zeta$ which
should be now halved.

\section{Mass spectrum}
\label{sec:mass}

Now we are ready to calculate particle mass-ratios in SGM $L\to\infty$ limit
using connection with spin chain (\ref{htf}). First we must numericaly
calculate mass
gaps of spin chain for finite $N$ and $h$. Then we must make continuum limit,
i.e.~take $N\to\infty$ keeping $L=Na$ and $\tilde{\mu}$ fixed (obviously $a\to
0$ and from (\ref{tmh}) $h\to 0$). Finally we should make $L\to\infty$,
i.e.~$\tilde{\mu}\to\infty$ (see (\ref{mhn})) limit. In practice, it is
preferable
to do following \cite{HenSal,Hen,Hen2}; first take $N\to\infty$ with $h$ fixed
and afterwords extrapolate to $h\to 0$. A difference is that in latter case one
does $\tilde{\mu}\to\infty$ before $h\to 0$. This limits are performed
using BST extrapolation method \cite{BulSt,HenSchutz}.

We numericaly diagonalized Hamiltonian (\ref{htf}) for up to 16 sites using
Lanczos algorithm. But before doing numerics, one should maximally exploite
symmetries. The Hamiltonian (\ref{htf})
commutes with translation operator $T$ (given by (\ref{txz})) and with
charge conjugation operator $C$. So, we can break Hamiltonian (\ref{htf})
in blocks, each marked with eigenvalues of the operators $P=i\ln T$ and $C$
which can be $P_{k}=\frac{2\pi i}{N}k\bmod 2\pi$, (see (\ref{pxz})) and
$C=\pm 1$ (because $C^{2}=1$). We are interested in mass ratios, so we
only need zero-momentum sector. But, because ``true'' space
translations are generated by $T^{2}$ (or because we must subtract
``nonuniversal macroscopic momentum'' $\pi$, if you like it more) zero-momentum
sector is a union of $P=0$ and $P=\pi$ sectors. So we must diagonalize four
blocks which we will denote by $0^{+}$, $0^{-}$, $\pi^{+}$ and $\pi^{-}$.

We considered a number of values of coupling $-1\le\Delta <1$ (or $\sqrt{2\pi}
\ge\beta >0$, see (\ref{brg})). Starting from $\Delta=-1$ spectrum contains
five clearly isolated states: vacuum and second breather in $0^{+}$, first
breather in $0^{-}$, soliton in $\pi^{-}$ and antisoliton in $\pi^{+}$. All
other levels form ``continuum'', i.e.~they ``densely'' fill the region
between $\approx 2\times(\mbox{mass of first breather})$ and some $E_{max}$.
Soliton and antisoliton energies are not degenerate which is a consequence
of breaking $\tilde{Z}_{2}$ symmetry on the spin chain. Exactly at
$\Delta=-1$ we have \cite{foot2} $\tilde{m}_{B1}=\tilde{m}_{S}<\tilde{m}_{A}
<\tilde{m}_{B2}$. As we
increase $\Delta$ $\tilde{m}_{S}$, $\tilde{m}_{A}$ and $\tilde{m}_{B2}$
monotonicaly increase (relative to $\tilde{m}_{B1}$) where $\tilde{m}_{S}$ and
$\tilde{m}_{A}$ increase faster than $\tilde{m}_{B2}$ and at $\Delta\approx
-0.1$ disappear into the ``continuum'' (i.e.~$\tilde{m}_{S,A}>2\tilde{m}_{B1}$
), while $\tilde{m}_{B2}$ asimptoticaly aproach $2\tilde{m}_{B1}$. This was a
crude picture visible already from row data before extrapolation $N\to\infty$
and $h\to 0$, and it is expected from the DHN formula (\ref{dhn}). Observe that
the exact degeneracy of soliton and first breather masses at $\Delta=-1$ is
present in (\ref{dhn}).

In Figs. \ref{mg9}-\ref{mg1} we present numerical resuls for the scaled
gaps (scaling
functions of mass gaps) $\tilde{G}_{a}$, $a\in\{ S,A,B1,B2\}$ at $\Delta=
-0.9,-0.6,-0.1$. This is of course check of the scaling relation
(\ref{tmhg}). BST extrapolations $N\to\infty$ (with fixed $h$) of scaled gaps
for $h=0.8,0.5,0.3,0.2$ are given in Tables \ref{g9}-\ref{g1}. As expected
convergence is better for higher $\Delta$.

To make an extrapolation $h\to 0$ one should obtain results for smaller $h$,
at least $h\cong 0.1$. From Figs. \ref{mg9}-\ref{mg1} one can see that for
that one should diagonalize Hamiltonian with $N\ge 26$, which is too
demanding even for most powerfull machines today.

Finally, (partialy) extrapolated mass ratios
\bdm 
\tilde{r}_{a}(\Delta,h)=\lim_{\stackrel{N\to\infty}{h\;\mathrm{fixed}}}
\frac{\tilde{m}_{a}}{\tilde{m}_{B1}}=\lim_{\stackrel{N\to\infty}{h\;
\mathrm{fixed}}}\frac{\tilde{G}_{a}}{\tilde{G}_{B1}},\qquad a\in\{ S,A,B2\}
\edm
are given in Tables \ref{r10}-\ref{r1} together with the predictions from
DHN formula (\ref{dhn}) and Fujita et al.~formula (\ref{fum}). One can see
that our results confirm DHN and reject Fujita et al.

\section{UV (conformal) limit of particle states}
\label{sec:conf}

Let's now turn our attention to the opposite UV limit of our results for
the spin chain (\ref{htf}). We saw in Sec. \ref{sec:gauss} that it is obtained
when $\mu (\tilde{\mu})\to 0$. Using (\ref{mem}) and (\ref{tmh}) in the
continuum result (\ref{mch}) we obtain that the scaling relation for mass gaps
of spin chain should have form
\bdm 
\tilde{m}_{a}(\gamma,\tilde{\mu},h)=\zeta h^{\frac{2\pi}{3\pi+\gamma}}
\left[ 2\pi d_{a}\tilde{\mu}^{-\frac{2\pi}{3\pi+\gamma}}+\tilde{H}_{a}
(\gamma,\tilde{\mu})\right]
\edm
where we must now include proper normalization factor $\zeta$ for the spin
chain Hamiltonian. Because it doesn't depend on $h$ we can take it from
unperturbed XXZ spin chain (\ref{xxz}), where it is well known
\bdm 
\zeta=\frac{\pi\sin\gamma}{\gamma}
\edm
Before we plot reduced scaling functions $\tilde{H}_{a}(\gamma,\tilde{\mu})$
we must know scaling dimension $d_{a}$ of the corresponding state. On the
other hand, we can choose $d_{a}$ and see does it gives the right behaviour
of $\tilde{H}_{a}(\gamma,\tilde{\mu})$ when $\tilde{\mu}\to 0$ (which is the
same as for $H_{a}$ mentioned below (\ref{xcp})).

In Table \ref{klme} we have presented scaling dimensions of zero-momentum
particle
states of SGM as conjectured in \cite{KlMelz}. But our numerical results
clearly indicate that first and second breather ($B2$ and $B2$) have exactly
the same scaling dimensions. In Figs. \ref{mh9}-\ref{mh1} we show numeric
results for
reduced scaling functions, where we used for scaling dimensions values from
Table \ref{our}.

We can see in Figs. \ref{mh9}-\ref{mh1} that finite size effects are stronger
for $\Delta$
closer to $-1$ (where they are in fact logaritmic because of the appearance
of marginal operators), which is expected from \cite{WoEc}.

\section{Conclusion}

In this paper we use the XXZ spin chain in transverse field as a lattice
regularization of the sine-Gordon model proposed in \cite{KlMelz}. This
equivalence can be understood
e.g.\ from conformal perturbation theory. One of our goals was to calculate
by numerical analysis masses in the sine-Gordon theory. This is now of
interest because recent calculations based on numerical treatment of Bethe
Ansatz \cite{FuSeYa} and infinite momentum frame technique \cite{FuOg} are
in disagreement with previous approaches used in literature
\cite{DHN,Zam1,BeTh,Kor}. Our results are in agreement with
DHN formula contrary to previously mentioned papers. We stress that methods
used in this paper are independent of previous approaches to SGM (which were
criticized in \cite{FuOg,FuSeYa,FuHi}). We also analyse conformal
limit and find conformal dimensions of various states. We find that the
conformal dimension of second breather state disagrees with the conjecture
by \cite{KlMelz}. Our calculations for dimensions of other states agree with
those in \cite{KlMelz}.

\begin{figure}
\caption{Scaling functions $\tilde{G}_{a}(\beta,\mu)$ for the isolated gaps
of Hamiltonian (\ref{htf}) at $\Delta=-0.9$ (or $\beta^2=5.38$). A legend in
upper left figure applies to all figures in this article.}
\label{mg9}
\end{figure}

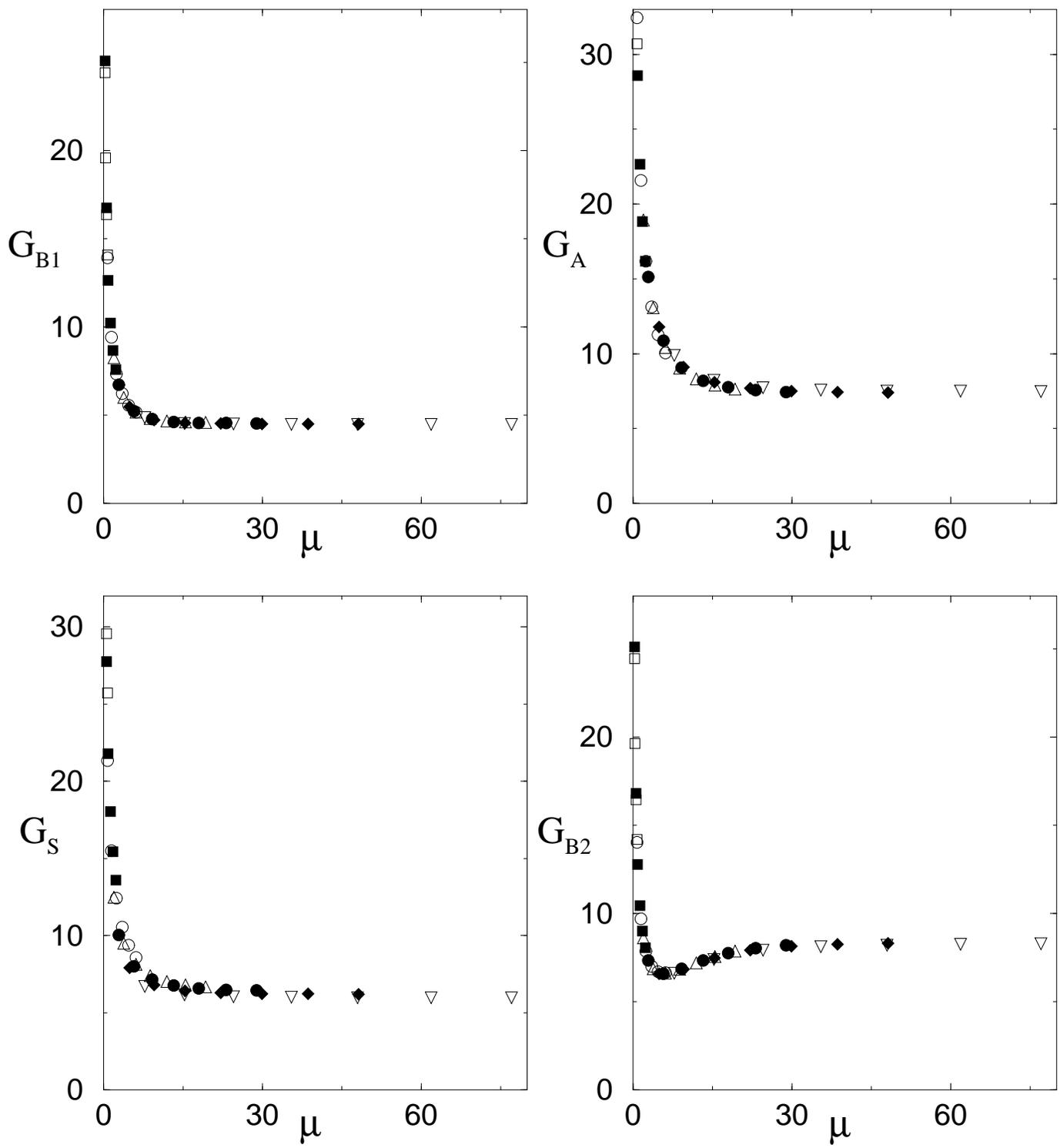
\begin{figure}
\caption{The same as Fig. \ref{mg9} but now for $\Delta=-0.6$ (or
$\beta^2=4.43$).}
\label{mg6}
\end{figure}

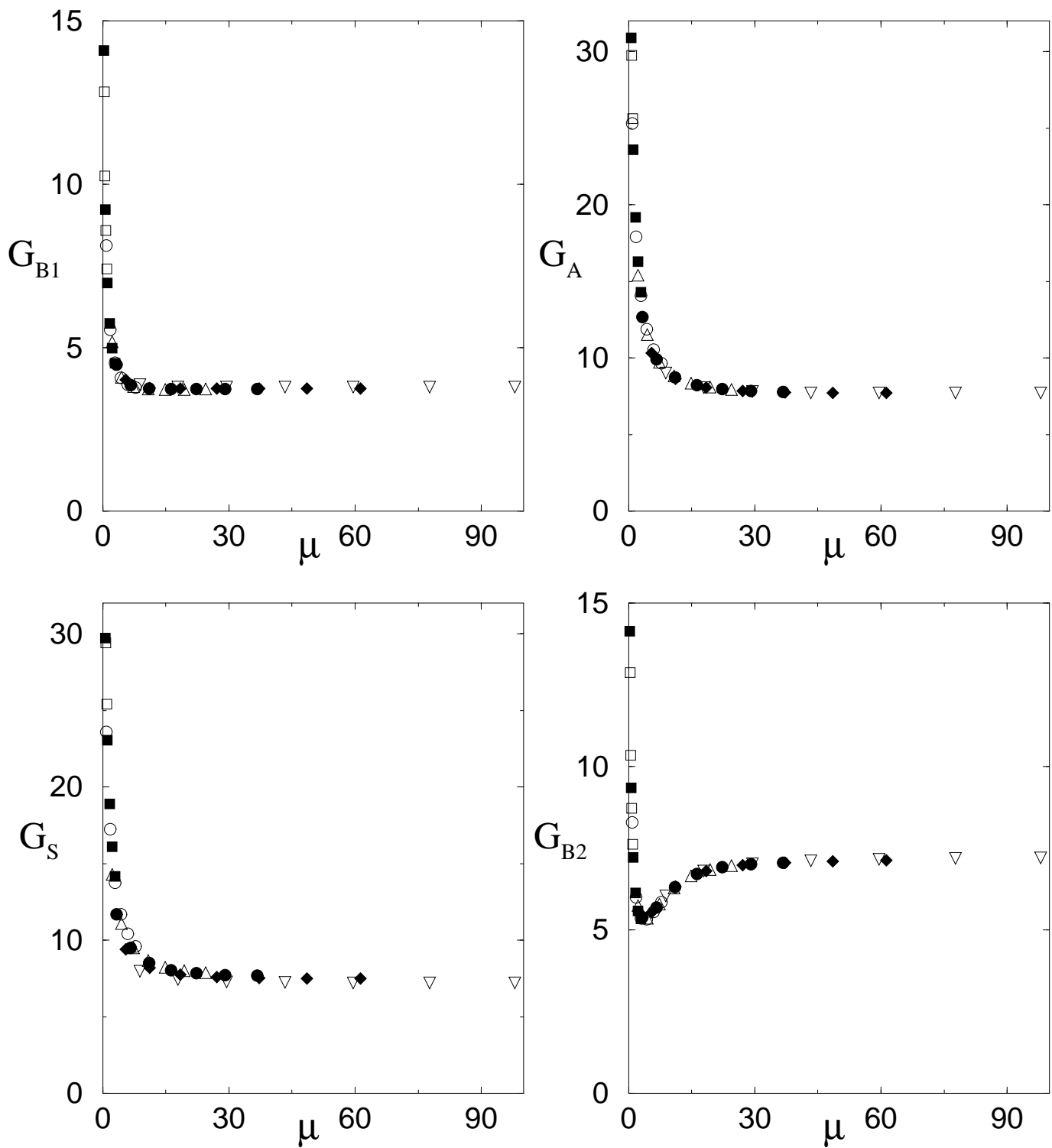
\begin{figure}
\caption{The same as Fig. \ref{mg9} but now for $\Delta=-0.1$ (or
$\beta^2=3.34$).}
\label{mg1}
\end{figure}

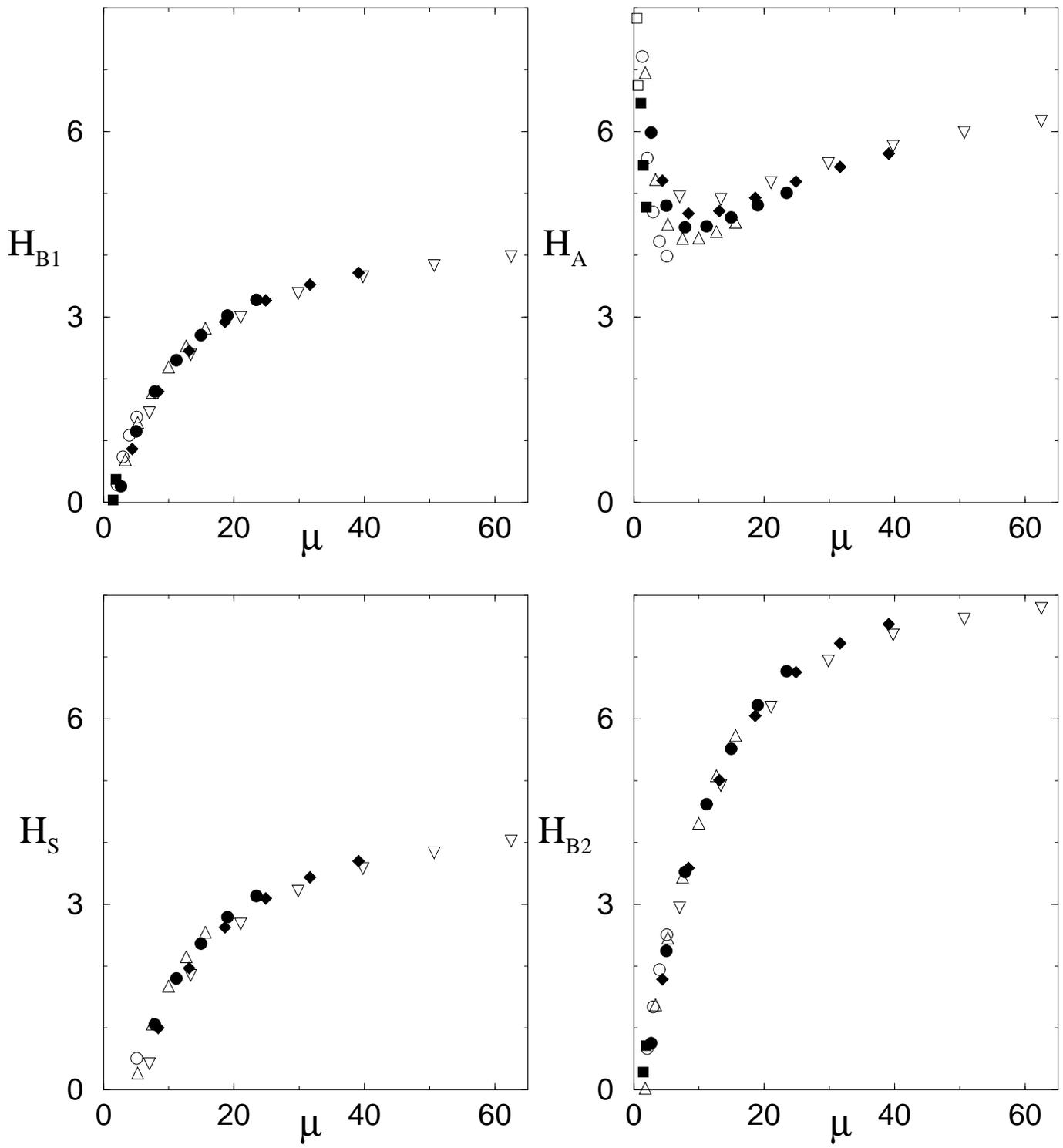
\begin{figure}
\caption{Reduced scaling functions $\tilde{H}_{a}(\beta,\mu)$ at $\Delta=-0.9$
(or $\beta^{2}=5.38$). A legend is the same as in Fig. \ref{mg9}.}
\label{mh9}
\end{figure}

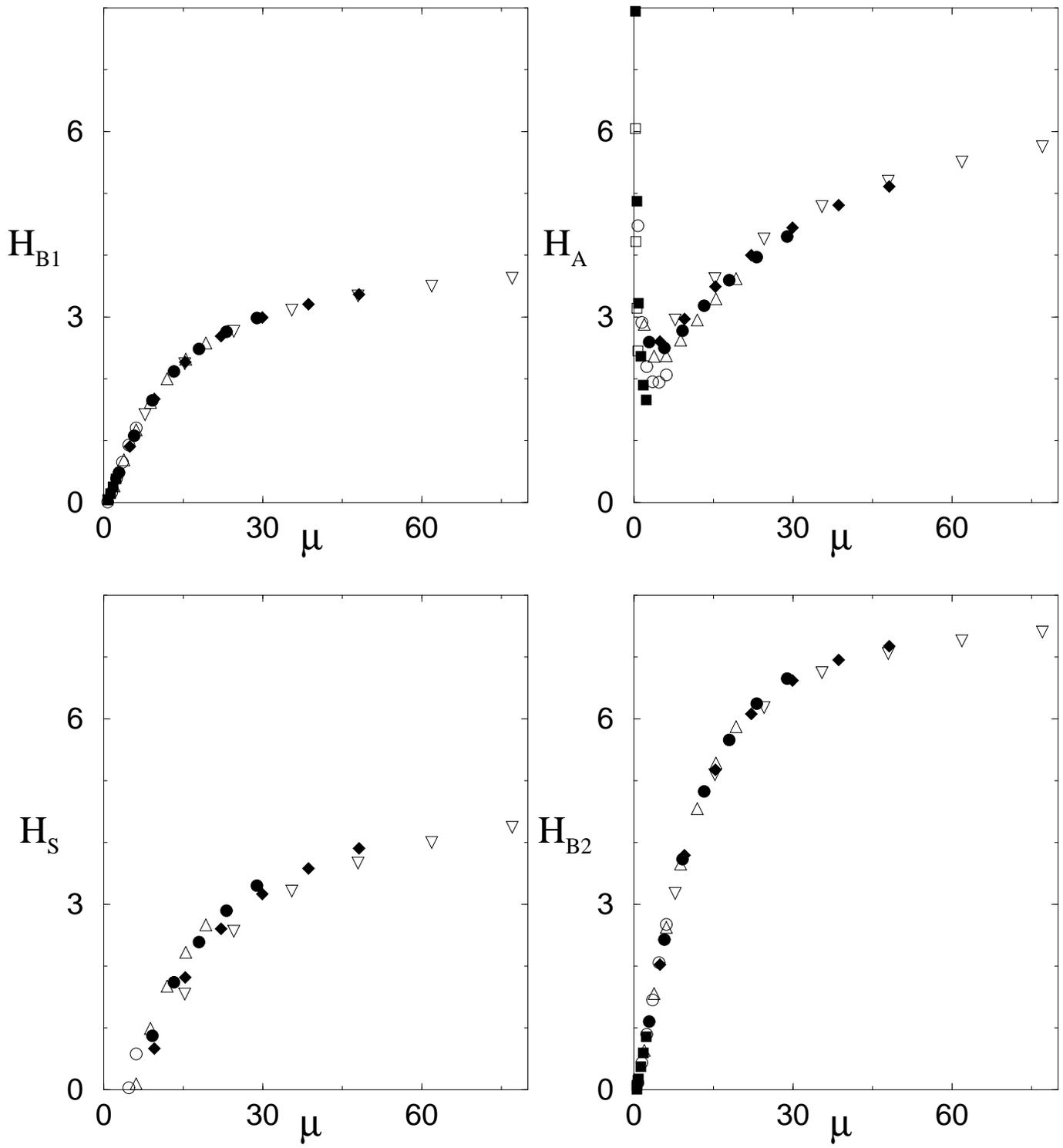
\begin{figure}
\caption{The same as Fig. \ref{mh9} but now for $\Delta=-0.6$ (or
$\beta^2=4.43$).}
\label{mh6}
\end{figure}

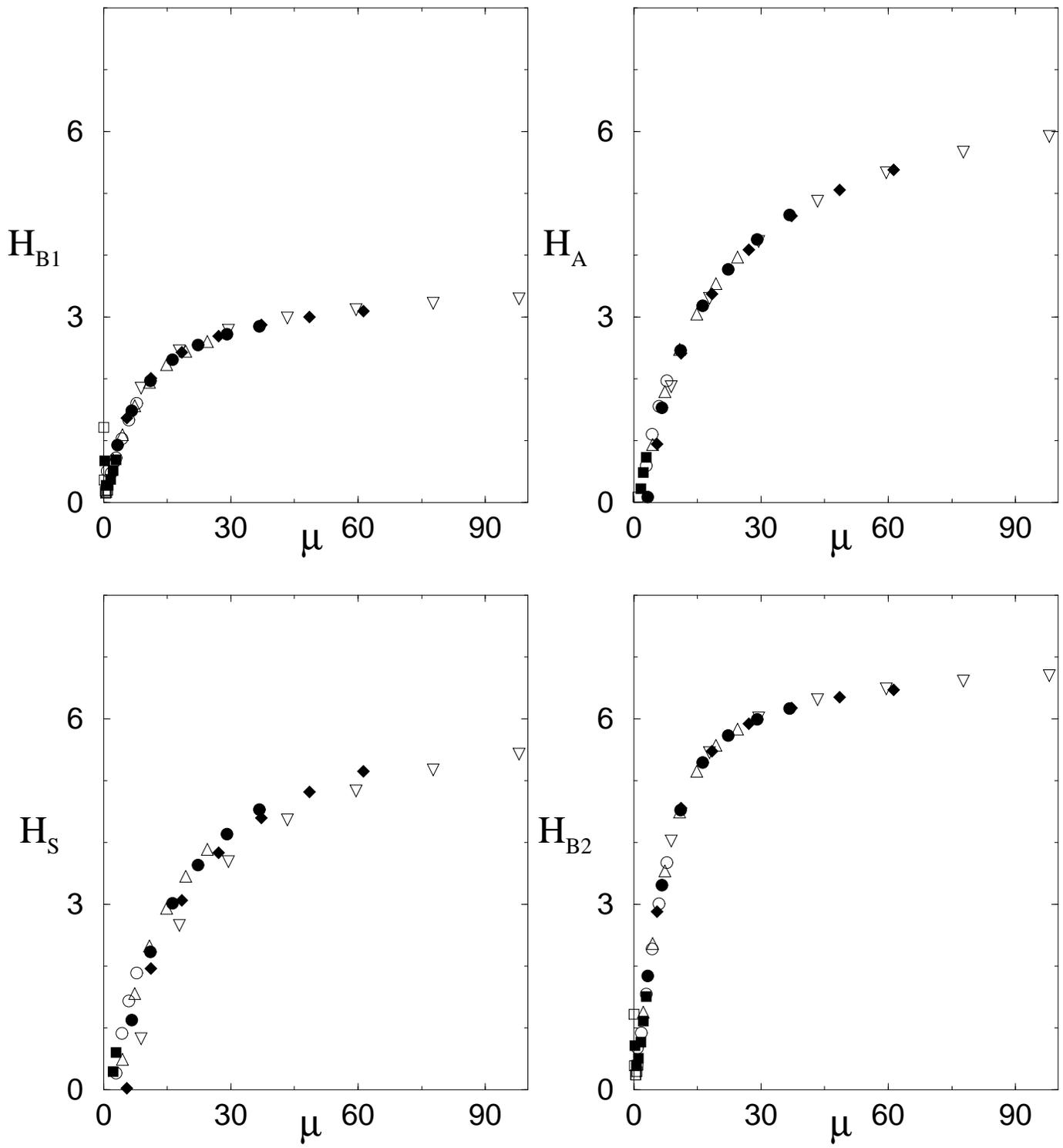
\begin{figure}
\caption{The same as Fig. \ref{mh9} but now for $\Delta=-0.1$ (or
$\beta^2=3.34$).}
\label{mh1}
\end{figure}

\begin{table}
\caption{Scaling dimensions of particle states in SGM as conjectured in
\protect\cite{KlMelz}.}
\begin{tabular}{lcc}
State & Operator & Scaling dimension \\
\tableline \rule[-2mm]{0mm}{6mm}
soliton & $V_{0,1}$ & $\frac{\pi}{\beta^{2}}$ \\ \rule[-2mm]{0mm}{6mm}
antisoliton & $V_{0,-1}$ & $\frac{\pi}{\beta^{2}}$ \\ 
\rule[-2mm]{0mm}{6mm}
p-th breather & $V_{p,0}^{((-)^{p})}$ & $\frac{p\beta^{2}}{4\pi}$ \\
\end{tabular}
\label{klme}
\end{table}


\begin{table}
\caption{Estimates for the scaled gaps
$\tilde{G}_{a}(\beta,\infty)$ as a function of $h$ at $\Delta=-0.9$
($\beta^{2}=5.38$). The numbers in brackets give the estimated uncertainty in
the last given digit.}
\begin{tabular}{lllll}
\multicolumn{1}{c}{$h$} & \multicolumn{1}{c}{$\tilde{G}_{B1}$} &
\multicolumn{1}{c}{$\tilde{G}_{S}$} & \multicolumn{1}{c}{$\tilde{G}_{A}$}
& \multicolumn{1}{c}{$\tilde{G}_{B2}$} \\ \hline
0.8 & 4.85922\,(5) & 5.2274\,(1) & 7.358\,(2) & 8.706\,(6) \\
0.5 & 4.9421\,(7) & 5.368\,(1) & 7.25\,(1) & 8.93\,(3) \\
0.3 & 5.012\,(6) & 5.49\,(1) & 7.10\,(5) & 9.2\,(1) \\
0.2 & 5.04\,(2) & 5.55\,(3) & 6.9\,(1) & 8.7\,(2) \\
\end{tabular}
\label{g9}
\end{table}

\begin{table}
\caption{The same as Table \ref{g9} but now for $\Delta=-0.6$
($\beta^{2}=4.43$).}
\begin{tabular}{lllll}
\multicolumn{1}{c}{$h$} & \multicolumn{1}{c}{$\tilde{G}_{B1}$} &
\multicolumn{1}{c}{$\tilde{G}_{S}$} & \multicolumn{1}{c}{$\tilde{G}_{A}$}
& \multicolumn{1}{c}{$\tilde{G}_{B2}$} \\ \hline
0.8 & 4.48354\,(1) & 5.9727\,(1) & 7.477\,(1) & 8.305\,(4) \\
0.5 & 4.51002\,(3) & 6.199\,(1) & 7.386\,(6) & 8.41\,(1) \\
0.3 & 4.537\,(1) & 6.38\,(1) & 7.28\,(3) & 8.49\,(5) \\
0.2 & 4.548\,(5) & 6.47\,(3) & 7.16\,(7) & 8.56\,(13) \\
\end{tabular}
\label{g6}
\end{table}


\begin{table}
\caption{The same as Table \ref{g9} but now for $\Delta=-0.1$
($\beta^{2}=3.34$).}
\begin{tabular}{lllll}
\multicolumn{1}{c}{$h$} & \multicolumn{1}{c}{$\tilde{G}_{B1}$} &
\multicolumn{1}{c}{$\tilde{G}_{S}$} & \multicolumn{1}{c}{$\tilde{G}_{A}$}
& \multicolumn{1}{c}{$\tilde{G}_{B2}$} \\ \hline
0.8 & 3.795834\,(2) & 7.21140\,(8) & 7.7036\,(2) & 7.261\,(5) \\
0.5 & 3.75549\,(3) & 7.483\,(1) & 7.715\,(2) & 7.21\,(1) \\
0.3 & 3.7372\,(3) & 7.63\,(1) & 7.73\,(1) & 7.16\,(1) \\
0.2 & 3.728\,(1) & 7.65\,(3) & 7.71\,(4) & 7.11\,(2) \\
\end{tabular}
\label{g1}
\end{table}

\begin{table}
\caption{Estimates for the mass gap ratios
$\tilde{r}_{a}(\Delta,h)$ as a function of $h$ at $\Delta=-1$
($\beta^{2}=2\pi$). We also added predictions obtained from (\ref{dhn}) (DHN)
and (\ref{fum}) (Fujita at al).}
\begin{tabular}{cllllll}
 & \multicolumn{4}{c}{h} & & \\ \cline{2-5}
\raisebox{1.5ex}[0pt]{$\tilde{r}_{a}$} & \multicolumn{1}{c}{0.8}
& \multicolumn{1}{c}{0.5} & \multicolumn{1}{c}{0.3} &
\multicolumn{1}{c}{0.2} & \raisebox{1.5ex}[0pt]{DHN} &
\raisebox{1.5ex}[0pt]{Fujita} \\ \hline
S & 1 & 1 & 1 & 1 & 1 & 0.877 \\
A & 1.4703\,(7) & 1.419\,(4) & 1.36\,(1) & 1.32\,(2) & 1 & 0.877 \\
B2 & 1.762\,(2) & 1.766\,(7) & 1.74\,(2) & 1.62\,(5) & 1.732 & 
\multicolumn{1}{c}{-} \\
\end{tabular} 
\label{r10}
\end{table}

\begin{table}
\caption{The same as Table \ref{r10} but now for $\Delta=-0.9$
($\beta^{2}=5.38$).}
\begin{tabular}{cllllll}
 & \multicolumn{4}{c}{h} & & \\ \cline{2-5}
\raisebox{1.5ex}[0pt]{$\tilde{r}_{a}$} & \multicolumn{1}{c}{0.8}
& \multicolumn{1}{c}{0.5} & \multicolumn{1}{c}{0.3} &
\multicolumn{1}{c}{0.2} & \raisebox{1.5ex}[0pt]{DHN} &
\raisebox{1.5ex}[0pt]{Fujita} \\ \hline
S & 1.07577\,(3) & 1.0862\,(2) & 1.095\,(3) & 1.101\,(7) & 1.205 & 1.018 \\
A & 1.5142\,(5) & 1.467\,(3) & 1.42\,(1) & 1.37\,(3) & 1.205 & 1.018 \\
B2 & 1.792\,(1) & 1.807\,(7) & 1.84\,(3) & 1.73\,(5) & 1.820 & 
\multicolumn{1}{c}{-} \\
\end{tabular}
\label{r9}
\end{table}

\begin{table}
\caption{The same as Table \ref{r10} but now for $\Delta=-0.6$
($\beta^{2}=4.43$).}
\begin{tabular}{cllllll}
 & \multicolumn{4}{c}{h} & & \\ \cline{2-5}
\raisebox{1.5ex}[0pt]{$\tilde{r}_{a}$} & \multicolumn{1}{c}{0.8}
& \multicolumn{1}{c}{0.5} & \multicolumn{1}{c}{0.3} &
\multicolumn{1}{c}{0.2} & \raisebox{1.5ex}[0pt]{DHN} &
\raisebox{1.5ex}[0pt]{Fujita} \\ \hline
S & 1.33214\,(3) & 1.3745\,(2) & 1.406\,(3) & 1.423\,(7) & 1.517 & 1.229 \\
A & 1.6677\,(2) & 1.638\,(1) & 1.605\,(8) & 1.57\,(2) & 1.517 & 1.229 \\
B2 & 1.8523\,(9) & 1.865\,(3) & 1.87\,(1) & 1.88\,(3) & 1.888 &
\multicolumn{1}{c}{-} \\
\end{tabular}
\label{r6}
\end{table}

\begin{table}
\caption{The same as Table \ref{r10} but now for $\Delta=-0.4$
($\beta^{2}=3.96$).}
\begin{tabular}{cllllll}
 & \multicolumn{4}{c}{h} & & \\ \cline{2-5}
\raisebox{1.5ex}[0pt]{$\tilde{r}_{a}$} & \multicolumn{1}{c}{0.8}
& \multicolumn{1}{c}{0.5} & \multicolumn{1}{c}{0.3} &
\multicolumn{1}{c}{0.2} & \raisebox{1.5ex}[0pt]{DHN} &
\raisebox{1.5ex}[0pt]{Fujita} \\ \hline
S & 1.53365\,(3) & 1.5970\,(2) & 1.639\,(3) & 1.654\,(8) & 1.724 & 1.367 \\ 
A & 1.7927\,(1) & 1.779\,(1) & 1.762\,(6) & 1.74\,(1) & 1.724 & 1.367 \\
B2 & 1.880\,(1) & 1.886\,(3) & 1.885\,(5) & 1.90\,(2) & 1.914 &
\multicolumn{1}{c}{-} \\
\end{tabular}
\label{r4}
\end{table}

\begin{table}
\caption{The same as Table \ref{r10} but now for $\Delta=-0.1$
($\beta^{2}=3.34$).}
\begin{tabular}{cllllll}
 & \multicolumn{4}{c}{h} & & \\ \cline{2-5}
\raisebox{1.5ex}[0pt]{$\tilde{r}_{a}$} & \multicolumn{1}{c}{0.8}
& \multicolumn{1}{c}{0.5} & \multicolumn{1}{c}{0.3} &
\multicolumn{1}{c}{0.2} & \raisebox{1.5ex}[0pt]{DHN} &
\raisebox{1.5ex}[0pt]{Fujita} \\ \hline
S & 1.89982\,(2) & 1.9926\,(3) & 2.042\,(4) & 2.052\,(9) & 2.096 & 1.612 \\ 
A & 2.02949\,(8) & 2.0543\,(7) & 2.068\,(3) & 2.07\,(1) & 2.096 & 1.612 \\
B2 & 1.913\,(1) & 1.920\,(3) & 1.916\,(4) & 1.907\,(7) & 1.942 &
\multicolumn{1}{c}{-} \\
\end{tabular}
\label{r1}
\end{table}

\begin{table}
\caption{Scaling dimensions of particle states in
SGM as conjectured from our numerical results.}
\begin{tabular}{lcc}
State & Operator & Scaling dimension \\
\hline \rule[-2mm]{0mm}{6mm}
soliton & $V_{0,1}$ & $\frac{\pi}{\beta^{2}}=\frac{1}{2}\left(
1-\frac{\gamma}{\pi}\right)^{-1}$ \\ \rule[-2mm]{0mm}{6mm}
antisoliton & $V_{0,-1}$ & $\frac{\pi}{\beta^{2}}=\frac{1}{2}\left(
1-\frac{\gamma}{\pi}\right)^{-1}$ \\
\rule[-2mm]{0mm}{6mm}
1-st breather & $V_{1,0}^{(-)}$ & $\frac{\beta^{2}}{4\pi}=\frac{1}{2}\left(
1-\frac{\gamma}{\pi}\right)$ \\
\rule[-2mm]{0mm}{6mm}
2-nd breather & $V_{1,0}^{(+)}$ & $\frac{\beta^{2}}{4\pi}=\frac{1}{2}\left(
1-\frac{\gamma}{\pi}\right)$ \\
\end{tabular}
\label{our}
\end{table}


\begin{references}
\bibitem{foot1}
  In fact, exactly that happens when one connects MTM to XYZ spin chain
  \cite{Lusch}.
\bibitem{foot2}
  We employ an obvious notation for mass gaps; $\tilde{m}_{S}$ for soliton,
  $\tilde{m}_{A}$ for antisoliton and $\tilde{m}_{Bn}$ for n-th breather
  where $n=1,2$.
\end{references}
\end{document}